\documentclass[prb,twocolumn,superscriptaddress, floatfix,nobibnotes]{revtex4}
\usepackage{amsmath,graphicx,epstopdf,color,bm,soul,upgreek, mathtools,siunitx,blindtext}
\newcommand{\add}[1]{\textcolor{blue}{\uline{#1}}}
\begin{document}
\renewcommand{\add}[1]{\textcolor{blue}{{#1}}}
\newcommand{\delete}[1]{\textcolor{red}{#1}}

\title{Topological Optical Waveguiding of Exciton-Polariton Condensates}

\author{J. Beierlein}
\email{johannes.beierlein@uni-wuerzburg.de}
\affiliation{Technische Physik, Wilhelm-Conrad-R\"ontgen-Research Center for Complex
Material Systems, and  W\"urzburg-Dresden Cluster of Excellence ct.qmat, Universit\"at W\"urzburg, Am Hubland, D-97074 W\"urzburg,
Germany}

\author{O. A. Egorov} 
\affiliation{Institute of Condensed Matter Theory and Optics,  Friedrich-Schiller-Universit\"at Jena, Max-Wien-Platz 1, D-07743 Jena, Germany}

\author{P. Gagel}%
\affiliation{Technische Physik, Wilhelm-Conrad-R\"ontgen-Research Center for Complex
Material Systems, and  W\"urzburg-Dresden Cluster of Excellence ct.qmat, Universit\"at W\"urzburg, Am Hubland, D-97074 W\"urzburg,
Germany}

\author{T.H. Harder}%
\affiliation{Technische Physik, Wilhelm-Conrad-R\"ontgen-Research Center for Complex
Material Systems, and  W\"urzburg-Dresden Cluster of Excellence ct.qmat, Universit\"at W\"urzburg, Am Hubland, D-97074 W\"urzburg,
Germany}

\author{A. Wolf}%
\affiliation{Technische Physik, Wilhelm-Conrad-R\"ontgen-Research Center for Complex
Material Systems, and  W\"urzburg-Dresden Cluster of Excellence ct.qmat, Universit\"at W\"urzburg, Am Hubland, D-97074 W\"urzburg,
Germany}

\author{M. Emmerling}%
\affiliation{Technische Physik, Wilhelm-Conrad-R\"ontgen-Research Center for Complex
Material Systems, and  W\"urzburg-Dresden Cluster of Excellence ct.qmat, Universit\"at W\"urzburg, Am Hubland, D-97074 W\"urzburg,
Germany}

\author{S. Betzold}%
\affiliation{Technische Physik, Wilhelm-Conrad-R\"ontgen-Research Center for Complex
Material Systems, and  W\"urzburg-Dresden Cluster of Excellence ct.qmat, Universit\"at W\"urzburg, Am Hubland, D-97074 W\"urzburg,
Germany}
\author{F. Jabeen}%
\affiliation{Technische Physik, Wilhelm-Conrad-R\"ontgen-Research Center for Complex
Material Systems, and  W\"urzburg-Dresden Cluster of Excellence ct.qmat, Universit\"at W\"urzburg, Am Hubland, D-97074 W\"urzburg,
Germany}


\author{L. Ma} 
\affiliation{Institute for Integrative Nanosciences Leibniz IFW Dresden, and W\"urzburg-Dresden Cluster of Excellence ct.qmat, D-01069 Dresden, German}

\author{S. H\"ofling}
\affiliation{Technische Physik, Wilhelm-Conrad-R\"ontgen-Research Center for Complex
Material Systems, and  W\"urzburg-Dresden Cluster of Excellence ct.qmat, Universit\"at W\"urzburg, Am Hubland, D-97074 W\"urzburg,
Germany}

\author{U. Peschel} 
\affiliation{Institute of Condensed Matter Theory and Optics, Friedrich-Schiller-Universit\"at Jena, Max-Wien-Platz 1, D-07743 Jena, Germany}

\author{S. Klembt}
\email{sebastian.klembt@physik.uni-wuerzburg.de}
\affiliation{Technische Physik, Wilhelm-Conrad-R\"ontgen-Research Center for Complex
Material Systems, and  W\"urzburg-Dresden Cluster of Excellence ct.qmat, Universit\"at W\"urzburg, Am Hubland, D-97074 W\"urzburg,
Germany}

\begin{abstract}
{One-dimensional models with topological non-trivial band structures are a simple and effective way to study novel and exciting concepts in topological photonics. In this work we are studying the propagation of light-matter quasi-particles, so called exciton-polaritons, in waveguide arrays. Specifically, we are investigating topological states at the interface between dimer chains, characterized by a non-zero winding number. In order to exercise precise control over the polariton propagation, we study non-resonant laser excitation as well as resonant excitation in transmission geometry. The results highlight a new platform for the study of quantum fluids of light and non-linear optical propagation effects in coupled semiconductor waveguides.}
\end{abstract}

\maketitle

\indent \textit{Introduction.} 
Topological physics \cite{Qi2011,Yan20212} and topological material \cite{Vergniory2019} have been attracting enormous interest over the last decade, promising exciting new effects and functionalities. Following a theoretical proposal by Haldane and Raghu \cite{Haldane2008} the realization of topological states of matter has also taken the field of photonics by storm. Emerging form the concept of Berry phases topological non-trivial states have been introduced in microwaves \cite{Wang2009}, Floquet \cite{Rechtsman2013} waveguides, Si photonics \cite{Hafezi2013}, metamaterials \cite{Mousavi2015} and many more \cite{Kane2013,Shindou2013}. In recent years, a rising interest has been noticeable in studying the interplay of topology and optical non-linearities. Here, topological lasers have been proposed \cite{Harari2018} and realized \cite{Bahari2017,Bandres2018,Dikopoltsev2021} in varying laser geometries. A system that is naturally endowed with a sizable nonlinear response are microcavity exciton-polaritons, resulting from the strong light-matter coupling of quantum well excitons with a photonic cavity mode. By deterministic coupling of microresonators forming polariton lattices \cite{Jacqmin2014,Winkler2015,Baboux2016}, Su-Schrieffer-Heeger (SSH) systems \cite{ St-Jean2017,Harder2021,Dusel2021,Pieczarka2021} as well as a 2D Chern insulators \cite{Klembt2018} have been realized. In addition, polaritons have be readily confined in waveguides \cite{Wertz2010,Walker2013} allowing for the formation and study of guided quantum fluids of light \cite{Wertz2012,Carusotto2013,Rozas2020,Beierlein2021a,Beierlein2021}.\\  Here, we uniquely combine an exciton-polariton platform with a patterned lattice potential landscape implementing a topological defect waveguide, effectively combining strong interactions with topologically non-trivial band structures. For this work, we have implemented a SSH model with coupled polariton waveguides. In order to achieve a sizable and controllable coupling - and staggered coupling for the SSH Hamiltonian - we employ and sophisticated etch-and-overgrowth method \cite{El-Daif2016,Winkler2017}, implementing the aforementioned coupling. 

In brief, the celebrated SSH model \cite{Su1979,Asboth2016}, originally introduced to describe the dynamics of non interacting spinless electrons in a 1D dimer chain,   is the simplest topological system that can support protected edge and defect states. The underlying chiral symmetry is responsible for the appearance of these topological edge and defect states in the middle of a band gap (zero-energy states) and protects them from perturbations. The principle of bulk-boundary correspondence states that the existence of the topological edge states in the lattice is predetermined by the topological invariant of the bulk, known as the winding number.
The SSH system is characterized by a dimerized coupling with two sites per unit cell, parameterized by different intra- (t) and inter-cell (t') hopping coefficients. In the tight-binding descriptions the system Hamiltonian can be written as

\begin{multline}
\widehat{H}=t \sum_{m=1}^{N}(|m,B\rangle \, \langle m,A| +h.c.) \\\ + t'\sum_{m=1}^{N-1} (|m+1,A\rangle \, \langle m,B| +h.c.),   
\end{multline}

where N denotes the number of a unit cell, A and B are the sites in the unit cell, and t and t' describe the hopping amplitudes, respectively. Calculating the winding number \cite{Cardano2017} one finds that a weakly bound waveguide as shown in Fig.~\ref{fig1}(a) $(t>t')$ leads to a topological defect state and a bulk winding number of $W = 1$.\cite{Blanco2016}\\
While SSH waveguides have been implemented in waveguide structures such as Si-photonics, floquet-type waveguides and plasmonic structures \cite{Blanco2016,Bleckmann2017,Fedorova2019,Kremer2021}, a demonstration involving exciton-polariton quantum fluids of light has been elusive so far.
Here, we are investigating two different polaritonic SSH waveguide system, allowing for complementary experiments into topological propagation in topological waveguides. Firstly, we performed experiments under non-resonant laser excitation in a GaAs QW-based cavity \cite{Wertz2010}, facilitating long-range polariton propagation. Furthermore, we are using an InGaAs QW-based sample that by careful backside polishing allows for deterministic resonant laser excitation of the gap and bulk modes of the SSH waveguides. This way polaritons can the deterministically injected into a mode of choice without having to rely on intricate relaxation mechanisms, opening the door to a wide range of new and exciting ways to study exciton-polariton propagation in topological systems.

\section{\label{sec:sample}Sample Fabrication and methods}

The two sample designed and investigated for this work consists of 37 bottom and 32 top Al\textsubscript{0.2}Ga\textsubscript{0.8}As/AlAs mirror pairs. Sample A has two stacks of four GaAs quantum wells (QWs) each with a width of \SI{7}{\nano\meter} and an exciton energy of $E\textsubscript{X} = \SI{1.614}{\electronvolt}$, as active material. Sample B utilizes one stack of three In\textsubscript{0.05}Ga\textsubscript{0.95}As QWs with a width of \SI{12}{\nano\meter} and a resulting exciton energy of  $E\textsubscript{X} = \SI{1.483}{\electronvolt}$ which were embedded in the structure. This structure has the advantage that the substrate and mirrors are fully transparent to the resulting polariton wavelengths, allowing for resonant excitation experiments in sample B. All QW stacks were positioned in the antinodes of the electric field  created by the $\lambda$/2-cavity and $\lambda$-cavity, respectively.

\begin{figure}[htb!]
\centering
\includegraphics[width=0.9\linewidth]{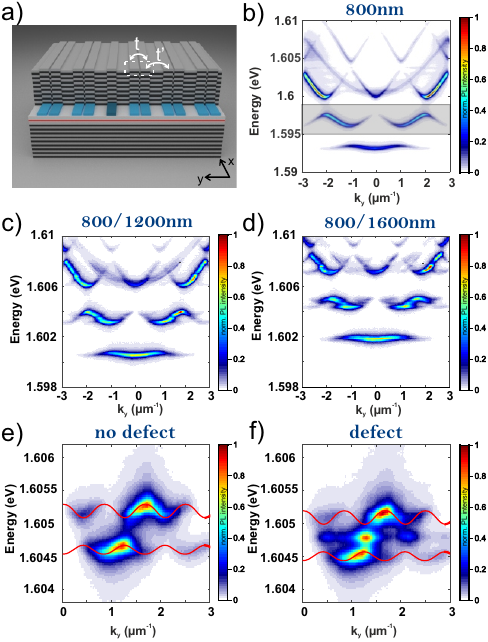} 

\caption{(a) Rendered schematic of the samples fabricated for this work. The stacks of quantum wells are illustrated in red while the created photonic potential in the cavity layer is highlighted in blue. The domain boundary is indicated by the dark blue.(b) Momentum-resolved dispersion in the linear regime of a waveguide array with homogeneous coupling between all waveguides, realized by a constant distance between the waveguides of \SI{800}{\nano\meter}. The P-band shows higher coupling due to its orbital nature (grey box). (c,d) Linear momentum-resolved dispersion's increasingly staggered waveguide arrays including a topological defect. A clear gap in the P-band is visible, as well as a defect state for 800/1600\,nm (d). (e,f) Zoomed in dispersion of a 800/1600\,nm lattice exciting only the staggered area without a defect (e) and including the easily visible topological defect (f). The red lines are the results of a bandstructure calculation highlighting the topological bandgap hosting a topological SSH defect in f). All dispersions were measured at $k_x =$\,\SI{0}{\per\micro\meter}.}
\label{fig1}
\end{figure}

Both samples have been grown by the so-called etch-and-overgrowth method\cite{El-Daif2016,Winkler2017,Beierlein2021}, where the growth of the microcavity is interrupted after the bottom distributed Bragg reflector (DBR) as well as the cavities has been grown. After that the sample is transfered out of the molecular beam epitaxy chamber and a combination of e-beam lithography and wet etching is used to impring the potential landscape of choice (SSH waveguides here) onto the sample. The photonic potential is created by etching into the cavity layer, therefore, reducing the effective cavity length in the desired area and creating a local photonic blueshift. The photonic potential can be precisely tuned by adjusting the etching depth and were adjusted to  $\sim$\,\SI{6.7}{\milli\electronvolt} and  $\sim$\,\SI{7.9}{\milli\electronvolt}, respectively, confining the S- and P-modes in the structures. Lastly, the sample is reintroduced into the growth chamber and the top DBR is grown onto the sample. To illustrate the process, a rendered schematic of the fabricated sample is presented in Fig.~\ref{fig1}(a). For the entire work, the $x$-direction denotes the direction along the waveguides, while $y$ denotes the direction perpendicular to the waveguides.  The top DBR is partially removed for clarity.  
Rabi-splittings of \SI{11.5}{\milli\electronvolt} and \SI{4.2}{\milli\electronvolt}, respectively have been determined via white light reflectivity measurements for sample A and B, displaying typically expected coupling strengths for GaAs and InGaAs QW polariton microcavities. By investigating polaritons with a highly photonic fraction the quality factors of both wavers were determined to  $\sim 7.500 $ and $\sim 5.000$, respectively.

The experimental results presented in the paper are predominantly obtained using real space and angle-resolved photoluminescence spectroscopy.
The samples presented in section~\ref{sec:non_res}. was located inside a liquid flow cryostat capable of keeping the sample at   $T$ = \SI{4.2}{\kelvin}, while for the results of section~\ref{sec:res}. a cryostat, where the sample is optically accessible from two directions, was used. Here, the sample typically reaches a temperature of  $T$ = \SI{10}{\kelvin}. In both cases a tuneable continuous wave Ti:sapphire laser was used to excite the samples. To adjust the excitation parameters, size and angle of the excitation beam, different lenses were used to shape the beam on the back focal plane of the objective. More detailed description is given in the respective sections of this paper. The optical setup is configured to allow angle or spatial resolution of the emission, which then is energetically resolved inside a Czerny-Turner monochromator and detected on Peltier-cooled charge-coupled device camera. Additionally, by using a motorized lens, the fully energy resolved real- and Fourier-space can be accessed using tomography techniques.

\begin{figure*} 
\centering
\includegraphics[width=\textwidth]{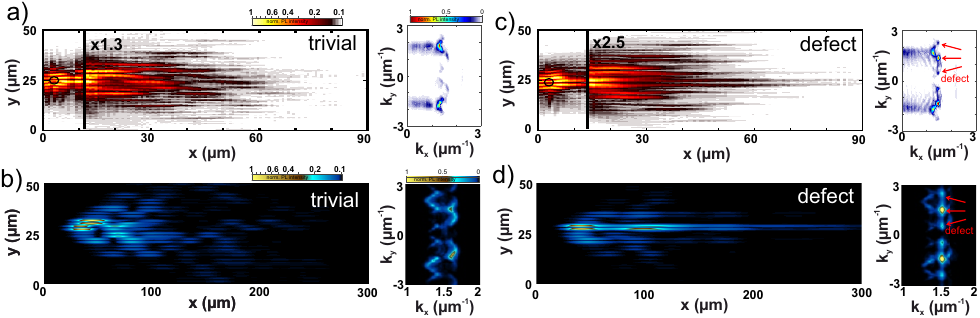} 
\caption{(a,c) Energy-resolved real space image of a P-band exciton polariton condensate propagating along a staggered waveguide lattice, having alternating gaps of \SI{800}{\nano\meter} and \SI{1200}{\nano\meter}, without (a) and with a topological defect (c). The right panels show the momentum-resolved dispersion, of the propagating condensate along the x-axis ($k_x >0$) , showing the original modal shape of the propagation inside the trivial P-mode (a) and additionally the topological defect mode (b). (b,d) Theoretical simulation of a condensate propagating along a waveguide array without (b) and with a defect (d) of topological non-trivial nature. The red arrows in (c,d) highlight the characteristic defect signature in k-space, symmetric in $k_y$.}
\label{fig2}
\end{figure*}

\section{\label{exp}Experimental Results}
\subsection{\label{sec:non_res}Non-resonant excitation}

In Fig.~\ref{fig1}(a)  a schematic of a staggered waveguide array (blue),  which is patterned into the cavity layer is shown. The direction along the waveguide propagation is denoted as the x-axis while the staggered waveguide pattern is oriented along the y-axis. Apart from the staggered waveguide distance, the lattice also features a defect or domain boundary, which in this case is defined by two consecutive large distances between the waveguides. In order to study the effect of staggering the distances between the waveguide one can start at the momentum-resolved dispersion of a homogeneous waveguide array, where the distance between all waveguides is \SI{800}{\nano\meter}. All waveguides presented in this work are \SI{2}{\micro\meter} wide. The corresponding dispersion is shown in Fig.~\ref{fig1}(b). Here, the coupling between the modes of the waveguide gives rise to a band structure, where the bandwidth of each band corresponds to the coupling of named modes. The coupling of the P-modes (grey box) is substantially higher than for the S-modes, due to its modal distribution and the larger overlap of the wavefunction \cite{Beierlein2021}.When increasing the staggering and therefore the distance between every second waveguide, it becomes obvious, that the P-band starts to split (see Fig.~\ref{fig1}(c-d)).

When the staggered lattice is interrupted by a domain boundary SSH defect, a mode located inside the gap arises. Since the opening of the gap is of topological nature, the state is protected from scattering into the bulk. The gap state is located primarily on the defect, still depending on the gap width and linewidth of the state and bands. Fig.~\ref{fig1}(e) shows the P-band of a staggered lattice with \SI{800}{\nano\meter} and \SI{1600}{\nano\meter} distance excited on the uniformly staggered region. A clear gap at \SI{1.605}{\electronvolt} is visible. The dispersion of the same lattice excited on a domain boundary is shown in Fig.~\ref{fig1}(f). Here, the emergence of a gap state is visible. 

Further, we investigate the propagation of a topological state in the polariton condensate regime along a waveguide array. By exciting the sample with a small spot, one can not only reach the polariton condensate regime, with low excitation power but also creates a local potential, originating from the repulsive interactions between carriers and the compound particles of the excitons. As a consequence of this potential, polaritons are rapidly expelled out of the excitation region and the condensation takes place at finite k-values and therefore finite group velocities.  In the suppl. mat. additional details on the condensation, the geometrical properties and dispersions along and across the waveguides can be found. In Fig.~\ref{fig2}(a) a polariton condensate, which is propagating along a waveguide array with an energy of \SI{1.6055}{\electronvolt} is depicted. The image shows an array with \SI{800}{\nano\meter} and \SI{1200}{\nano\meter} distances, respectively. As the condensate propagates, the population decreases due to the inherent radiative loss of the cavity. The excitation spot has a width of \SI{2}{\micro\meter} and is highlighted by a white circle. The grey lines are indicators for different ND filter to adjust for the decaying intensity. The coupling including a slight asymmetry is visible, which is to be expected in a staggered lattice. This is dependent on the excitation position in relation to the unit cell\cite{Bleckmann2017}. The inset shows the propagating condensate in the Fourier space at the same energy. The propagating mode has the same shape as shown in Fig.~\ref{fig1}(d) but with a wavevector $k_x \neq ~$\SI{0}{\per\micro\meter}.

When exciting the waveguide array on the topological defect, the propagation and thus the emission pattern changes substantially (see Fig.~\ref{fig2}(c)). A pronounced emission of polaritons propagating along the SSH defect is visible and coupling to the bulk region is significantly  reduced. The gap state is exponentially localized having a density probability in the bulk on every second waveguide\cite{Blanco2016}. Since the defect is located in the P-band, the real space emission shows two distinct lines, which are the lobes of the P-mode both originating from the same (single) defect waveguide. The modal shape of the defect is again shown in the inset.
While the emission in the defect mode is dominant, there is propagation in the bulk mode. This can be explained by the size of the gap in comparison to the linewidth. Here, the gap state can still couple slightly to the bulk.

To underpin these experimental results, we performed numerical simulation of the respective exciton-polariton dynamics within the mean-field model based on two coupled Schrödinger equations for the intracavity photonic field and coherent excitons in quantum wells. For the details see suppl. mat. and Ref. \cite{Carusotto2013,Beierlein2021a,Beierlein2021}.  
Within the modeling, we skip details of condensation dynamics and focus on the propagation of polaritons in the SSH waveguide array. Therefore, the initial polaritons with a appropriate frequency and momenta are launched by a localized coherent pump beam. Figures~\ref{fig2}b) and d) show the results of numerical simulations of polariton propagation dynamics in the potential landscape closely related to the experimental configuration for a homogeneously staggered lattice and the lattice with the topological defect, respectively. The appropriate choice of the momentum of the launched beams as well as the size and exact position of the spot allows for a direct excitation of the desired waveguide modes (see suppl. materials). Similar to the experimental results, the exciton-polaritons launched into the S-mode do not couple to neighbouring waveguides, thus keeping their energy in the excitation guides. Typical discrete diffraction dynamics occurs after excitation of a higher waveguide mode (P-mode) in a homogeneous SSH waveguide array, as it is shown in Fig.~\ref{fig2}b). The respective two-dimensional Fourier transformation of the polariton profiles shows the typical (for SSH lattice) dispersion relation consisting of two Bloch bands separated by a gap (see the inset in Fig.~\ref{fig2}b)). In contrast, by excitation of the SSH waveguide array with a defect the respective topological state can be excited as it is shown in Fig.~\ref{fig2}d). This  topological defect mode occurs in the middle of the gap between two Bloch-bands (see the inset in Fig.~\ref{fig2}d)). The k-space measurements in  Fig.~\ref{fig2}b) and d) precisely show the same signature in experiment and theory highlighted by red arrows for $k_y>0$.

\subsection{\label{sec:res}Resonant Excitation}

\begin{figure}  [tb!]
\centering
\includegraphics[width=0.9\linewidth]{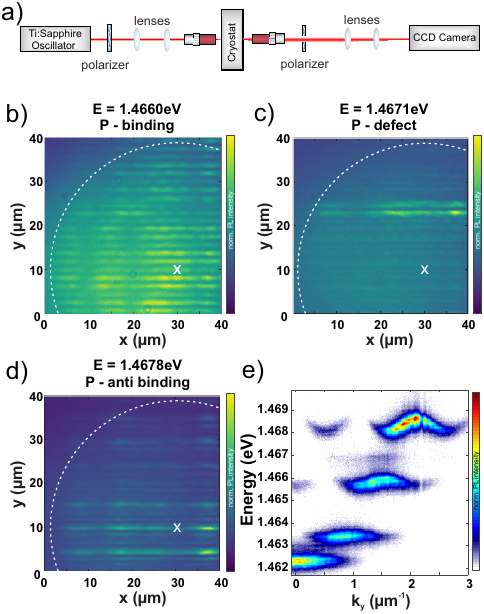} 
\caption{ a) Schematic of the beam path for the transmission setup, including beam shaping and signal detection lenses.   b)-d) Polariton photoluminescence measurement in transmission geometry with resonant excitation under $k_{x,y}= $\,\SI{0}{\per\micro\meter} of the respective P-binding, topological defect and P-anti binding modes in real space. The intensity distribution is a convolution of the excitation spot which is centered at x\,=\,\SI{30}{\micro\meter}, y\,=\,\SI{10}{\micro\meter} with a FWHM of $\,\sim$\,\SI{56}{\micro\meter} and the modal distribution affected by the energy gradient due to the growth. The excitation spot is highlighted with a white X and the the FWHM with a dashed line, respectively.
e) Polariton waveguide dispersion, measured in reflection geometry and under non-resonant excitation in confinement direction $k_y$ (for $k_x=$\,\SI{0}{\per\um}) , where the P-modes are visible between $\sim$\,\SI{1.466}{\electronvolt} and $\sim$\,\SI{1.468}{\electronvolt}.}

\label{fig3}
\end{figure}

In order to do more in depth study of the topological state, we have introduced a setup allowing for resonant laser excitation of the polaritonic modes using a transmission path in the setup. Here, the microcavity hosting InGaAs QWs (sample B) is used, since the DBRs as well as the GaAs substrate are transparent at the resonance wavelength of the resulting polaritonic modes. A simplified schematic of the beam path can be seen in Fig.~\ref{fig3}a). In the injection path, the beam can be focused to different sizes and positions onto the back focal plane of the injection objective to adjust the beam size and direction. Additionally, elliptical lenses where used later to adjust the shape of the excitation. 
The excitation laser is filtered from the detection path via cross polarization. The transmission geometry is highly advantageous, since scattering of the excitation laser is greatly reduced when compared to reflection measurements, allowing to efficiently filter out the laser and study the resonantly excited polaritons \cite{Harder2020}. 
The microcavity used for this section was grown on a single side polished wafer, which, due to the roughness of the backside surface, prevents unperturbed optical backside access to the sample. Therefore we have carefully polished the backside of the sample using a diamond powder milling technique. More information is presented in the supplement.

 Due to the parabolic nature of the dispersion of a microcavity photon, the k-values of the subsequent modes are increasing towards higher energies. To isolate and resonantly excite individual single modes, the excitation beam is focused on the backfocal plane ($k_x$ and $k_y\sim$\,\SI{0}{\per\um}) of the injection objective. Consequently the sample is illuminated in a wide area with a spot size of approximately $\sim$\,\SI{56}{\micro\meter}. Fig.~\ref{fig3}b)-d) show a waveguide array in real space with \SI{200}{\nano\meter} and \SI{1200}{\nano\meter} distances for the energies \SI{1.4660}{\electronvolt}, \SI{1.4671}{\electronvolt} and  \SI{1.4678}{\electronvolt}, respectively. These represent the bulk P-binding, P-anti binding as well as the topological defect in the P mode. 

Fig.~\ref{fig3}b) depicts the binding P-mode extending virtually over the entire waveguide array. The emission profile is mainly dominated by the laser excitation shape. In combination, a small influence of the (photon to exciton) detuning gradient of the cavity which is naturally induced during growth is visible. For this sample the gradient is \SI{37}{\micro\electronvolt\per\micro\meter} along the y-direction. When now shifting the laser excitation energy to the topological gap around $E_{topo}\sim$\, \SI{1.4672}{\electronvolt} the strongly localized defect mode becomes clearly visible in Fig.~\ref{fig3}c). It is located in the center waveguide and presents with the dumbbell shape, typical for the P-mode. For higher energies the anti-binding P-mode can be detected in Fig.~\ref{fig3}d).  Additionally, growth related defect lines along the y-axis can be seen in Fig.~\ref{fig3}b). These can alter the transmission. Any difference in the mode shape along the x-direction would indicate irregularities in the etching or overgrowth process. In Fig.~\ref{fig3}e) the momentum-resolved dispersion of the waveguide array is depicted for positive wavevectors in $k_y$ (for $k_x=$\,\SI{0}{\per\um}) for reference. The reference dispersion was measured in reflectivity geometry and under non-resonant excitation. The size of the gap in the S-band as well as the P-band is  \SI{14.5}{\milli\electronvolt} an \SI{20.5}{\milli\electronvolt}, respectively. The topological defect of the P-mode is visible inside the gap. The S-band however, should be regarded as only slightly coupled molecules as the coupling strength of the S-mode for \SI{1200}{\nano\meter} is way smaller than the linewidth of \SI{450}{\micro\electronvolt}. 

\begin{figure} [tb!]
\centering
\includegraphics[width=\linewidth]{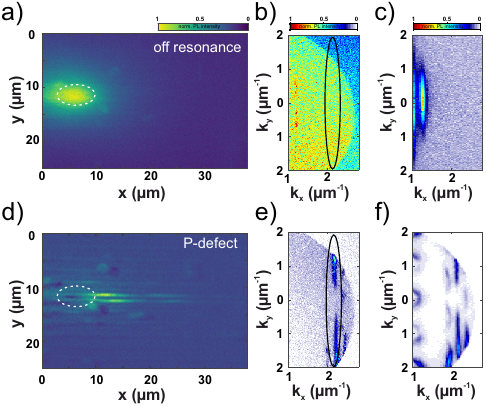} 
\caption{ Transmission measured in real space a) and momentum space b) off resonance. The excitation spot is indicated with a white dashed line. In momentum space, the laser excitation is at a finite k value and shaped as an ellipse as indicated by the black circle. The energy is E\,=\,\SI{1.4641}{\electronvolt}. c) Momentum-resolved dispersion for the same energy measured prior in reflectance under non resonant excitation. d) Real space image of the resonantly excited topological defect mode at an energy of E\,=\,\SI{1.472}{\electronvolt}. e) Corresponding momentum resolved image of the topological defect mode. f) Non-resonant excited dispersion at the same energy as d) showing the rest of the P-band.}
\label{fig4}
\end{figure}

In pursuit of a more comprehensive understanding of the physics governing the propagation of topological states, it is crucial to launch a polariton wavepacket resonantly within a topological waveguide mode at a finite energy and wavevector. Therefore, the laser is shaped to an elliptical spot, off center of the back focal plane. This corresponds to a 90 degree turned elliptical spot on the sample surface, which is reaching the sample under a finite angle ($\sim$ 20\,degree). To demonstrate the potential of this approach, the real space transmission, momentum-resolved transmission and previously measured momentum-resolved reflection is shown in Fig.~\ref{fig4}. In Fig.~\ref{fig4}a) the transmission off resonance at an energy of \SI{1.4641}{\electronvolt} and a wavevector $k_x$ = \SI{2.1}{\per\micro\meter}is depicted. The excitation spot, is indicated with a white dashed ellipse in the real space image and a black dashed ellipse in the k space image. As no state is available the transmission in Fig.~\ref{fig4}b)is mainly scattered light. The s-band can already be seen at $k_x$ = \SI{1.3}{\per\micro\meter} for this energy in Fig.~\ref{fig4}b),c) .
Due to the parabolic dispersion of the microcavity polariton along the waveguide direction, one can address the states shown in Fig.~\ref{fig3}e) only at higher energies for $k_x\neq$\,\SI{0}{\per\um}), respectively. 
Subsequently, the energy is  raised to \SI{1.472}{\electronvolt}, were the topological defect mode with non-zero momentum comes into resonance with the excitation laser. The defect waveguide mode is clearly visible in Fig.~\ref{fig4}d) and compares well with the previous measurements in Fig.~\ref{fig3}c). The dispersion measured in transmission (Fig.~\ref{fig4}e)) is in good agreement with the previously measured dispersion in reflection geometry in Fig.~\ref{fig4}f).
The proximity in momentum space of the binding and anti-binding P-band in the reflected dispersion and the suppressed emission in transmission shows the excitation selectivity possible with this approach. The possibility to resonantly launch polariton wavepackets in an arbitrary waveguide potential opens up new possibility in the study of quantum fluids of light, their interactions, non-linearities and topology, e.g. in the context of Thouless pumping \cite{Lohse2016}, in lattice potentials.

\section{Conclusion}

In summary we have shown topological waveguiding of exciton-polaritons in SSH waveguide arrays. The momentum-resolved dispersion of waveguide arrays featuring topological band gaps and gap states were shown for two different material systems. The propagation of exciton polariton condensates in topological trivial and non trivial states have been observed and underlined by numerical simulations. Additionally, deterministically controlled resonant excitation of a topological polariton gap state at finite wavevectors was realized for the first time. The underlying methodology and approach has been described. This work shows highly controllable access to excite specific bands of complex polaritonic systems and pans the way for investigations of the interplay of topological non-trivial band structures with highly nonlinear quantum fluids of light. 

\begin{acknowledgments}
The W\"urzburg and Jena group acknowledge financial support within the DFG projects, PE 523/18-1 and KL3124/2-1 . 
The W\"urzburg Group acknowledges financial support by the DFG under Germany’s Excellence Strategy - EXC2147 “ct.qmat” (Project No. 390 858 490) and is grateful for support by the state of Bavaria.

\newpage

\end{acknowledgments}

\end{document}


\title{Supplemental Material \\ Topological Optical Waveguiding of Exciton-Polariton Condensates}

\author{J. Beierlein}
\email{johannes.beierlein@uni-wuerzburg.de}
\affiliation{Technische Physik, Wilhelm-Conrad-R\"ontgen-Research Center for Complex
Material Systems, and  W\"urzburg-Dresden Cluster of Excellence ct.qmat, Universit\"at W\"urzburg, Am Hubland, D-97074 W\"urzburg,
Germany}

\author{O. A. Egorov} 
\affiliation{Institute of Condensed Matter Theory and Optics,  Friedrich-Schiller-Universit\"at Jena, Max-Wien-Platz 1, D-07743 Jena, Germany}

\author{P. Gagel}%
\affiliation{Technische Physik, Wilhelm-Conrad-R\"ontgen-Research Center for Complex
Material Systems, and  W\"urzburg-Dresden Cluster of Excellence ct.qmat, Universit\"at W\"urzburg, Am Hubland, D-97074 W\"urzburg,
Germany}

\author{T.H. Harder}%
\affiliation{Technische Physik, Wilhelm-Conrad-R\"ontgen-Research Center for Complex
Material Systems, and  W\"urzburg-Dresden Cluster of Excellence ct.qmat, Universit\"at W\"urzburg, Am Hubland, D-97074 W\"urzburg,
Germany}

\author{A. Wolf}%
\affiliation{Technische Physik, Wilhelm-Conrad-R\"ontgen-Research Center for Complex
Material Systems, and  W\"urzburg-Dresden Cluster of Excellence ct.qmat, Universit\"at W\"urzburg, Am Hubland, D-97074 W\"urzburg,
Germany}

\author{M. Emmerling}%
\affiliation{Technische Physik, Wilhelm-Conrad-R\"ontgen-Research Center for Complex
Material Systems, and  W\"urzburg-Dresden Cluster of Excellence ct.qmat, Universit\"at W\"urzburg, Am Hubland, D-97074 W\"urzburg,
Germany}

\author{S. Betzold}%
\affiliation{Technische Physik, Wilhelm-Conrad-R\"ontgen-Research Center for Complex
Material Systems, and  W\"urzburg-Dresden Cluster of Excellence ct.qmat, Universit\"at W\"urzburg, Am Hubland, D-97074 W\"urzburg,
Germany}
\author{F. Jabeen}%
\affiliation{Technische Physik, Wilhelm-Conrad-R\"ontgen-Research Center for Complex
Material Systems, and  W\"urzburg-Dresden Cluster of Excellence ct.qmat, Universit\"at W\"urzburg, Am Hubland, D-97074 W\"urzburg,
Germany}


\author{L. Ma} 
\affiliation{Institute for Integrative Nanosciences Leibniz IFW Dresden, and W\"urzburg-Dresden Cluster of Excellence ct.qmat, D-01069 Dresden, German}

\author{S. H\"ofling}
\affiliation{Technische Physik, Wilhelm-Conrad-R\"ontgen-Research Center for Complex
Material Systems, and  W\"urzburg-Dresden Cluster of Excellence ct.qmat, Universit\"at W\"urzburg, Am Hubland, D-97074 W\"urzburg,
Germany}

\author{U. Peschel} 
\affiliation{Institute of Condensed Matter Theory and Optics, Friedrich-Schiller-Universit\"at Jena, Max-Wien-Platz 1, D-07743 Jena, Germany}

\author{S. Klembt}
\email{sebastian.klembt@physik.uni-wuerzburg.de}
\affiliation{Technische Physik, Wilhelm-Conrad-R\"ontgen-Research Center for Complex
Material Systems, and  W\"urzburg-Dresden Cluster of Excellence ct.qmat, Universit\"at W\"urzburg, Am Hubland, D-97074 W\"urzburg,
Germany}

\maketitle

\section{Numerical modeling of the polartion propagation}
We performed numerical calculations of exciton-polartion dynamics in semiconductor microcavity where the required waveguiding geometry can be accounted by inducing an appropriate potential for photons. Neglecting polarization effects one obtaines two coupled Schrödinger equations for the intracavity photonic field $\Psi _c$ and coherent excitons $\Psi _e$  \cite{Carusotto2013,Beierlein2021} giving as
\begin{align}
{\partial _t}{\Psi _c} - & \frac{{i\hbar }}{{2{m_c}}}\nabla _{x,y}^2{\Psi _c}  + iV(y){\Psi _c}   \nonumber \\ 
&+ \left( {{\gamma _c} - i{\Delta _c}} \right){\Psi _c} = i{\Omega _R}{\Psi _e} + E_\text{p}{e^{i{k_p}x}}
\end{align}
\begin{equation}
{\partial _t}{\Psi _e} - \frac{{i\hbar }}{{2{m_{ex}}}}\nabla _{x,y}^2{\Psi _e} + \left( {{\gamma _e} - i{\Delta _e}} \right){\Psi _e} = i{\Omega _R}{\Psi _e}
\end{equation}
The complex amplitudes $\Psi _c$ and $\Psi _e$ are obtained by averaging related creation or annihilation operators. The effective photon mass in the planar region is given by $m_c = 31.74 \times 10^{-6} m_e$ where  $m_e$  is the free electron mass. The effective mass of excitons $m_{ex} \approx 10^{5} m_e$. The coupling strength between intracavity photons and excitons $\Omega _R$ defines the Rabi splitting $2\hbar \Omega _R = 9.0 meV$. The parameters $ \Delta _{c,e} = \omega _p - \omega _{c,e} $ account for the frequency detunings of the operation frequency $\omega_p$ from the cavity $\omega_c$ and excitonic $\omega_e$ resonances, respectively. Then the exciton photon detunings of the device is given by $ \hbar \omega_c - \hbar \omega_e = \hbar \Delta_e - \hbar \Delta_c = -9.0 meV  $.  $ \hbar \gamma _c = \hbar \gamma _e = 0.02 meV$ are the cavity photon and exciton damping constants. An external photonic potential $V(x,y)$ defines the SSH waveguide geometry induced by structuring of the planar microcavity. In our modeling, a separate waveguide profile of the array is given by a super-Gauss  $V(y)=V_{0}\cdot(1-\text{exp}(-y^{50}/s^{50}))$ with the potential depth $\hbar V_0 = 8 meV$ and waveguide width $2s$.     

Since we are interested in the propagation dynamics of polaritons with a well-defined frequency an momenta  it is sufficient to consider  the case of coherent excitation by a pump beam $E_\text{p} (x,y)$ with a frequency $\omega_p$ and with a momentum $k_\text{p}$. In order to excite the P-band of the underling SSH structure we used the following parameters of the three pumping beams: $\hbar \omega_e = -4.8 meV$ and $k_x=1.5 \mu m^{-1}$. As well as $k_y= \SIlist{-1.5;0;1.5}{\per\um}$.

\newpage

\section{Sample preperation}

\begin{figure}[h!]
\centering
\includegraphics[width=0.9\linewidth]{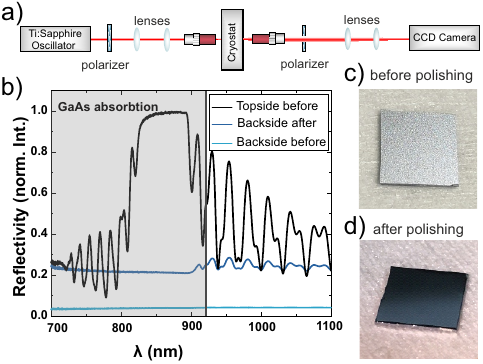}
\caption{a) Schematic of the beam path for the transmission setup, including beam shaping and signal detection lenses. b) Measured reflectivity of the sample's top and backside before and after backside polishing at room temperature. After polishing the tail of the DBR reflectivity through the absorption of the GaAs s observable. An image of the backside of the sample before c) and after d) backside polishing, respectively.}
\label{supfig1}
\end{figure}

The reflectance of the sample at room temperature is shown for Fig.~\ref{supfig1}b) for the topside before the polish, as well as the backside before and after. The topside shows, the expected stopband with Bragg minima. Before polishing no signal is reflected back directly from the sample, afterwards the tail of the stopband can be seen while lower wavelength are absorbed by the GaAs substrate. To emphasise the difference an image of the backside of the sample before and after polishing is depicted in Fig.~\ref{supfig1}b), respectively.

\section{Dispersion of polaritons below and above threshold}

\begin{figure}[tb!]
\centering
\includegraphics[width=0.9\linewidth]{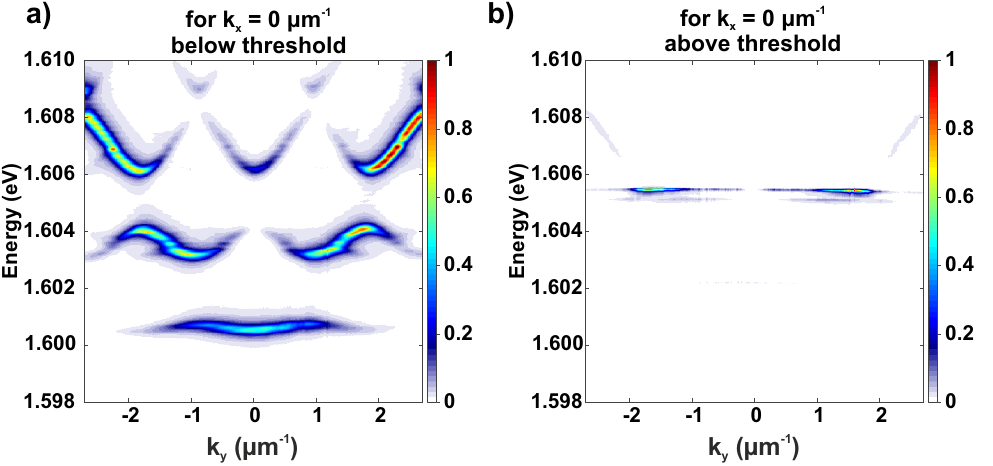} 
\caption{Polariton dispersion of a waveguide array with gap sizes \SI{800}{\nano\meter} and \SI{1200}{\nano\meter} along the $k_{y}$ direction, below a) and above threshold b), respectively. }
\label{supfig3}
\end{figure}

As the density in a polaritonic system rises, simulated scattering from the reservoir towards the ground states increases. The system undergoes a so-called polariton condensation. This is shown by a nonlinear increase in intensity, a linewidth drop and blue-shift of the system. Depending on the detuning of the polariton dispersion, higher modes can condensate before the ground mode. In Figure~\ref{supfig3} a polariton dispersion along the $k_{y}$ direction for a staggered waveguide array is shown below the condensation threshold. For increasing excitation powers and a small excitation spot the dispersion shifts to higher energies. For a waveguide system the polaritons condensate at a finite wavevector, also due to the repulsive potential created by the polariton reservoir at the excitation region. Additionally, due to the moderate negative detuning of the presented system the polaritons condense into the P-mode. The intensity detected at $k_{x}=$\,\SI{0}{\per\micro\meter} is now limited to the static portion of the polariton condensate, which indeed underwent a linewidth drop as well as a blueshift. Further details of the dispersion of a propagating polariton can be seen in Figure~\ref{supfig2}.  

\section{Dispersion of propagation polariton condensates}

\begin{figure} 
\centering
\includegraphics[width=0.9\linewidth]{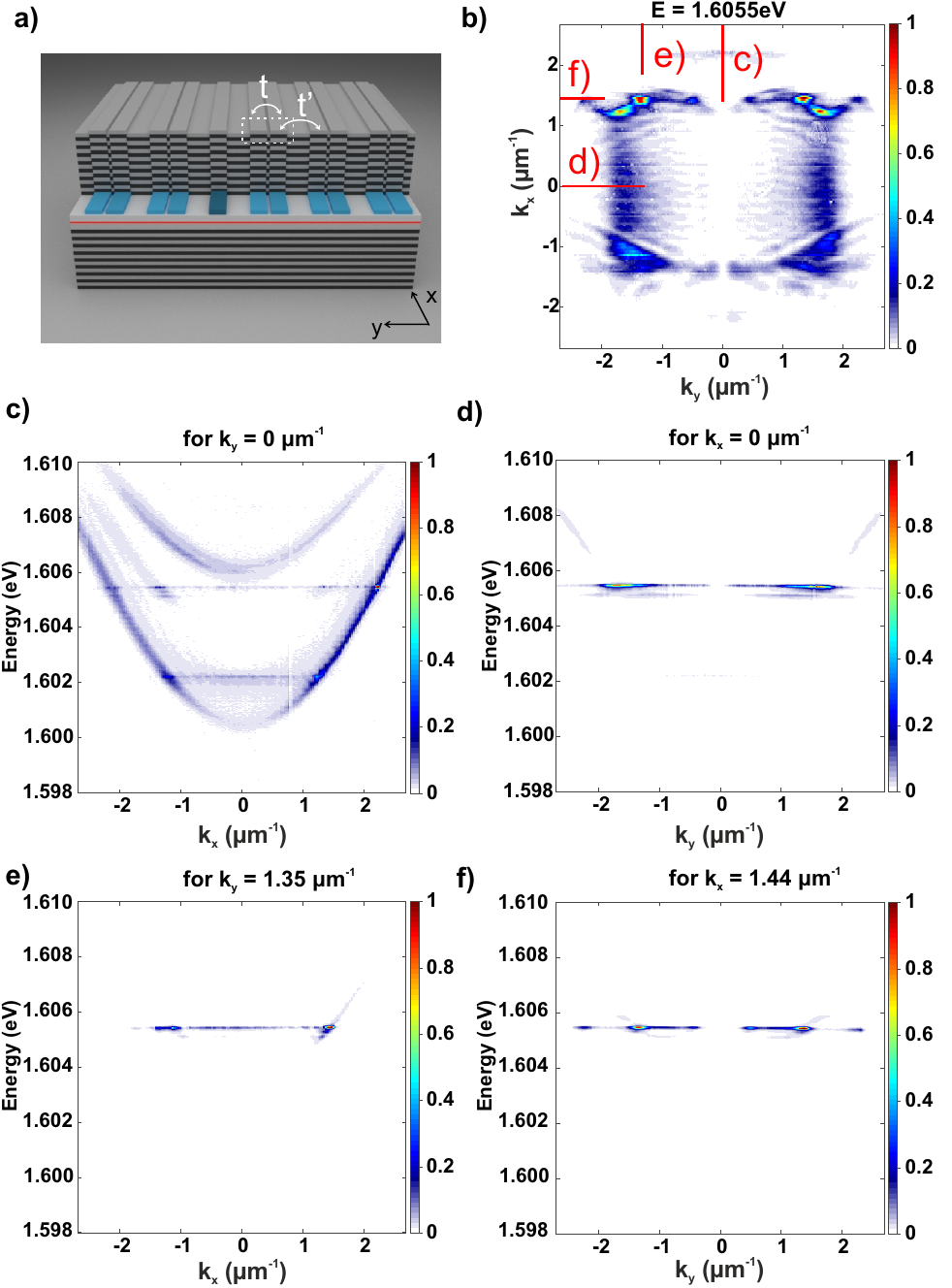} 
\caption{a) Rendered schematic of the samples fabricated for this work. The x-axis corresponds to the axis of propagation along the waveguides, while the staggered lattice is defined along the y-axis. b) Polariton waveguide dispersion in $k_{x,y}$ for a propagation condensate along the x-axis and an energy of E\,=\,\SI{1.6055}{\electronvolt}. c)-f) Various cuts through the $k_{x,y}$ dispersion. c) Momentum-resolved dispersion along the waveguides at $k_{y}=$\,\SI{0}{\per\micro\meter}, showing the parabolic nature of the dispersion. The P-mode parabola is strongly reduced in intensity as it has a singularity at $k_{y}=$\,\SI{0}{\per\micro\meter}. d) Momentum-resolved dispersion perpendicular to the waveguides at $k_{x}=$\,\SI{0}{\per\micro\meter}. e) Momentum-resolved dispersion along the waveguides at $k_{y}=$\,\SI{1.35}{\per\micro\meter}, showing the predominate propagation in positive x direction. f) Momentum-resolved dispersion perpendicular to the waveguides at $k_{x}=$\,\SI{1.44}{\per\micro\meter}, highlighting the modal shape of the condensate inside the defect state.}
\label{supfig2}
\end{figure}

After condensation in a waveguide the large majority of polaritons propagate at a finite wavevector along the unconfined direction of the waveguide. A schematic of the waveguide array investigated in this work is shown in Figure~\ref{supfig2}. The waveguides extend along the x-axis and are confined in the y-axis. The dispersion of the condensate launched in Figure 2b) of the main text over  $k_{x,y}$ is shown in full in Figure~\ref{supfig2}b) for E\,=\,\SI{1.6055}{\electronvolt}. Here, the information of the modal shape as well as relative wavevector can be seen. Additionally, we present various different slices through the dispersion in Figure~\ref{supfig2}c)-f). Here, we show for once the parabolic nature of the dispersion along $k_{x}$. The P-mode is not visible as the mode has a singularity in the field distribution at $k_{x}=$\,\SI{0}{\per\micro\meter}. The dispersion along $k_{y}$ depicts polaritons, which decayed before gaining finite momentum or were trapped in a local potential. The propagation condensate along the waveguide can be detected when looking at $k_{x}$ for $k_{y}=$\,\SI{1.35}{\per\micro\meter}. As the spot is located at the edge of the array, there is a finite portion propagating in negative $k_{x}$ direction. Finally the modal shape of the defect can be highlighted by plotting $k_{y}$ for $k_{x}=$\,\SI{1.44}{\per\micro\meter}. The modal shape is identical to mode shown in Figure~1~f) of the main text.

\section{Mode distribution of  propagating polariton condensates}

\begin{figure} 
\centering
\includegraphics[width=0.9\linewidth]{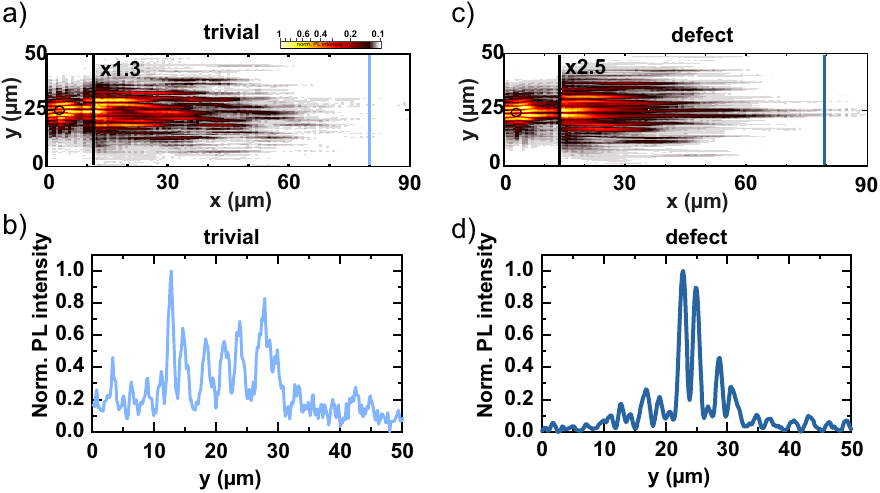} 
\caption{(a,c) Energy-resolved real space image of a P-band exciton polariton condensate propagating along a staggered waveguide lattice, having alternating gaps of \SI{800}{\nano\meter} and \SI{1200}{\nano\meter}, without a) and with a topological defect c). (b,d) Normalized intensity of the PL at $x=$\,\SI{80}{\micro\meter} for the topological trivial lattice b) and with a defect d) of topological non-trivial nature. The expected mode pattern at every second waveguide and the exponential decay of the mode density away form the center waveguide is observed.}
\label{supfig4}
\end{figure}

The real space distribution of a polariton condensate propagating a long a waveguide array is dependent on the mode it inherits during condensation. Figure~\ref{supfig4}(a,c) depicts the energy-resolved real space images of condensate propagating along a trivial a) and a waveguide array with a topological defect c). In order to underline the claim of the topological defect mode, the normalized intensities of the PL at a propagation distance of $x=$\,\SI{80}{\micro\meter} across the waveguides arrays are plotted in Figure~\ref{supfig4}(b,d) respectively. While in   Figure~\ref{supfig4}b) only the typical asymmetry created by the staggering and the limited excitation spot is visible. The condensate excited in the P-mode topological defect state shows its characteristic double lope shape and the expected mode pattern at every second waveguide and the exponential decay of the mode density away form the center waveguide.

